\documentclass[12pt]{article}
\usepackage{amsmath,amssymb,bm,graphicx}
\usepackage{graphicx}
\usepackage[dvipdfmx]{color} 
\begin{document}

\begin{flushright}
May 19, 2016
\end{flushright}

\vskip 0.5 truecm

\begin{center}
{\Large{\bf Dimensional regularization is generic\footnote{To be published in the Proceedings of  Conference on New Physics at the Large Hadron Collider,  Nanyang Technological University, Singapore,  29 February to 4 March,  2016.}}}
\end{center}
\vskip .5 truecm
\centerline{\bf  Kazuo Fujikawa}
\vskip .4 truecm
\centerline {\it RIKEN Nishina Center, Wako 351-0198, 
Japan}
\vskip 0.5 truecm

\makeatletter
\@addtoreset{equation}{section}
\def\theequation{\thesection.\arabic{equation}}
\makeatother

\begin{abstract}
The absence of the quadratic divergence in the Higgs sector of the Standard Model in the dimensional regularization is usually regarded to be an exceptional  property of a specific regularization.  To understand what is going on in the dimensional regularization, 
we illustrate how to reproduce the results of the dimensional regularization for the $\lambda\phi^{4}$ theory in the more conventional regularization such as the higher derivative regularization; the basic postulate involved is that the quadratically divergent induced mass, which is independent of the scale change of the physical mass, is kinematical and unphysical. This is consistent with the derivation of the Callan-Symanzik equation, which is a comparison of two theories with slightly different masses,  for the $\lambda\phi^{4}$ theory without encountering the quadratic divergence.  We thus suggest that the dimensional regularization is generic in a bottom-up approach starting with a successful low-energy theory. We also define a modified version of the mass independent renormalization for a scalar field which leads to the homogeneous renormalization group equation. Implications of the present analysis on the Standard Model at high energies and the presence or absence of SUSY at LHC energies are briefly discussed. 

\end{abstract}

\section{Introduction}

The quadratic divergence of the Higgs mass plays an important role in the precise re-evaluation  of the Standard Model and possible physics beyond it~\cite{weinberg,susskind}. This issue is also related to the notion of "naturalness", and also to the presence or 
absence of SUSY at LHC.  We discuss this issue of quadratic divergence mainly on the basis of Ref.~\cite{fujikawa} which was written before LHC but physics at LHC in mind, and now with more emphasis on the dimensional regularization.  We concentrate on the problem of quadratic divergence from a point of view of regularization, namely, a point of view that the {\em dimensional regularization is generic and physical} in a bottom-up approach starting with a successful low-energy theory.  

We start with an analysis of a typical quadratic divergence appearing in the $\lambda \phi^{4}$ theory, which is the only model we discuss in the present paper.
The Lagrangian is given by 
\begin{eqnarray}
{\cal L}=\frac{1}{2}\partial_{\mu}\phi(x)\partial^{\mu}\phi(x)-\frac{1}{2}
m_{0}^{2}\phi(x)^{2}-\frac{1}{4!}\lambda_{0}\phi(x)^{4}
\end{eqnarray}
which gives the  one-loop mass correction (after Wick rotation) 
\begin{eqnarray} \label{1.2}
\frac{\lambda_{0}}{2}\int\frac{d^{4}k}{(2\pi)^{4}}\frac{1}{k^{2}+m_{0}^{2}}
&=&\frac{\lambda_{0}}{32\pi^{2}}\int_{0}^{M^{2}}
dk^{2}\frac{k^{2}}{k^{2}+m_{0}^{2}}\nonumber\\
&=&\frac{\lambda_{0}}{32\pi^{2}}[M^{2}-m_{0}^{2}\ln\frac{M^{2}+m_{0}^{2}}{m_{0}^{2}}].
\end{eqnarray}
Thus the (renormalized) two-point effective potential is
\begin{eqnarray}
\Gamma_{2}(k, m_{0})=k^{2}+m_{0}^{2}+\frac{\lambda_{0}}{32\pi^{2}}[M^{2}-m_{0}^{2}\ln\frac{M^{2}+m_{0}^{2}}{m_{0}^{2}}]
\end{eqnarray}
since the wave function renormalization factor $Z=1$ to this order.
The conventional {\em multiplicative renormalization} suggests the replacement 
\begin{eqnarray}
m_{0}^{2}=Z_{m}m^{2}
\end{eqnarray}
with 
\begin{eqnarray}
Z_{m}=1+\frac{\lambda_{0}}{32\pi^{2}}\ln\frac{M^{2}+m_{0}^{2}}{\mu^{2}}.
\end{eqnarray}
We then have
\begin{eqnarray}
\Gamma_{2}(k, m_{0})&=&k^{2}+m^{2}+\frac{\lambda_{0}}{32\pi^{2}}[M^{2}-m_{0}^{2}\ln\frac{\mu^{2}}{m_{0}^{2}}]+O(\lambda_{0}^{2})\nonumber\\
&=&k^{2}+m^{2}\left(1-\frac{\lambda}{32\pi^{2}}\ln\frac{\mu^{2}}{m^{2}}\right)+\frac{\lambda_{0}}{32\pi^{2}}M^{2}+O(\lambda_{0}^{2}).
\end{eqnarray}
This quantity needs to be finite if the multiplicative renormalization
works. One thus needs to take care of the quadratic divergence term.
It is important to notice 
 the absence of the term
\begin{eqnarray}
 M^{2}\ln \frac{M^{2}}{m_{0}^{2}}
 \end{eqnarray}
 which is later shown to be the characteristic property of the quadratic divergence, and it is also the basis of our argument to the effect that the quadratic divergence is kinematical and unphysical. 

One may understand this divergence in (1.6) as:\\
(i) This leads to  the fine-tuning 
problem. It is ``unnatural'' that the Higgs particle stays 125 GeV.\\
(ii) One may subtract the quadratic divergence altogether so that one treats  the scalar mass by multiplicative renormalization. This is the scheme which supersymmetry(SUSY) and also dimensional regularization enforce.

The application of the dimensional regularization (or the simple subtraction) is generally called "{\em unnatural}",
since there appears to be no symmetry or dynamical reason to remove the enormously large induced mass. 
In contrast, SUSY scheme is usually called "{\em natural}"
since the basic supersymmetry ensures the multiplicative renormalization of the scalar mass just as the fermion mass.
The purpose of the present paper is to provide a ``physical reason''
for the subtraction of quadratic divergences by the dimensional regularization. 
\\

\noindent{\bf  SUSY and quadratic divergence:}\\

We first briefly recall the renormalization property of SUSY. 
The simplest model of supersymmetry in 4-dimensional space-time is the Wess-Zumino model~\cite{wess-zumino}.
\begin{eqnarray}
{\cal L}&=&i\partial_{n}\bar{\psi}\bar{\sigma}^{n}\psi +A^{\star}\Box A + F^{\star}F\nonumber\\
&&+[m(AF-\frac{1}{2}\psi\psi)+g(AAF-\psi\psi A) + h.c.]
\end{eqnarray}
which gives rise to the $A^{4}$ coupling if one integrates out the auxiliary
field $F$.  In this model, it is known that the {\em Non-renormalization theorem} holds:
The Wess-Zumino model is made finite by a uniform (logarithmically divergent) wave function renormalization
without even the finite renormalization of $m$ and $g$ when {\em renormalized at the vanishing momenta}.

All order proof of this fact has been given in the 
component formulation by J. Iliopoulos and B. Zumino~\cite{iliopoulos}, which used the higher derivative regularization.  A simplified proof using 
the superfield formulation was given by the present author and W. Lang~\cite{fujikawa1}, which emphasized the renormalization at vanishing momenta.  The superfield formulation introduced by Salam and Strathdee~\cite{salam} greatly simplifies the analysis when the  formulation is applicable.

The quadratic divergence is thus controlled by SUSY as a cancellation of contributions of fermions and bosons. 
But no simple regularization of {\em gauge invariant} SUSY models is known at this moment since:\\
(i) No superfield formulation for general supersymmetric gauge theory.\\
(ii) Higher derivative regularization of supersymmetric gauge theory is not known and that higher derivative regularization does not work for one-loop diagrams.\\
(iii) Dimensional regularization has complications in supersymmetry.\\
(iv) Lattice regularization has difficulties with the Leibniz rule among others.\\

Those complications may be just of technical
nature or may be of fundamental nature.  Even if SUSY is discovered at LHC, there are several technical issues to be resolved.\\

\noindent{\bf Dimensional regularization and quadratic divergence:}\\

We next recall the essence of the dimensional regularization.
The basic idea of the dimensional regularization is to extend the space-time dimensionality slightly away from $D=4$, and then take the limit $D=4$ after the 
actual calculation.  This idea was introduced by 
G. 't Hooft and M. Veltman~\cite{t hooft1}, 
C.G. Bollini and J.J. Giambiagi~\cite{bollini}, and J.F. Ashmore~\cite{ashmore}, and nicely summarized by J. Collins~\cite{collins}.

We analyze a simple example in \eqref{1.2}:
\begin{eqnarray}
\frac{\lambda_{0}}{2}\int \frac{d^{4}k}{(2\pi)^{4}} \frac{1}{k^{2}+m_{0}^{2}}\Rightarrow \frac{\lambda_{0}}{2}\int \frac{d^{D}k}{(2\pi)^{D}}
 \frac{1}{k^{2}+m_{0}^{2}}
\end{eqnarray}
and 
\begin{eqnarray}
\int d^{D}k \frac{1}{k^{2}+m_{0}^{2}}&=&\int d^{D}k\int_{0}^{\infty} ds e^{-s(k^{2}+m_{0}^{2})} \nonumber\\
&=&\int_{0}^{\infty}ds (\frac{\pi}{s})^{D/2}e^{-sm_{0}^{2}} \nonumber\\
&=&\pi^{D/2}(m_{0}^{2})^{D/2-1}\Gamma(1-D/2)
\end{eqnarray}
where
\begin{eqnarray}
x\Gamma(x)=\Gamma(1+x), \ \ \ \ \ \Gamma(1)=1 .
\end{eqnarray}
Near $\epsilon=2-D/2 \simeq 0$, we use
\begin{eqnarray}
A^{-\epsilon}&\simeq& 1-\epsilon \ln A, \nonumber\\
\Gamma(\epsilon-1)&=&\frac{1}{\epsilon(\epsilon-1)}\Gamma(1+\epsilon)\nonumber\\
&\simeq&\frac{1-\epsilon\gamma}{\epsilon(\epsilon-1)}\nonumber\\
&=&-\frac{1}{\epsilon}+\gamma-1
\end{eqnarray}
with $\gamma=-\Gamma^{\prime}(1)$.
We thus have
\begin{eqnarray}
\frac{\lambda_{0}}{2}\int \frac{d^{4}k}{(2\pi)^{4}} \frac{1}{k^{2}+m_{0}^{2}}
&\simeq&
\frac{(\lambda_{0}\mu^{-2\epsilon})}{32\pi^{2}}m_{0}^{2}[-\frac{1}{\epsilon}+\gamma-1
-\ln(\frac{\pi\mu^{2}}{4m_{0}^{2}})].
\end{eqnarray}
The correspondence with the momentum cut-off is 
\begin{eqnarray}
\frac{1}{\epsilon}=\frac{1}{2-D/2}\ \leftrightarrow \ \ln\frac{M^{2}}{\mu^{2}}
\end{eqnarray}
and the important feature is that {\em we do not encounter the quadratic divergence} in the dimensional regularization. Note also that the combination
$\lambda_{0}\mu^{-2\epsilon}$ is dimensionless in $\lambda_{0}\phi^{4}$ theory even away from $D=4$.

To my knowledge, the absence of the quadratic divergence in dimensional regularization was first emphasized by S. Weinberg who mentioned, "Also, any
quadratic divergences are always an artifact of
the cutoff procedure; they do not appear if we use
dimensional regularization."~\cite{weinberg}, when the fine tuning problem associated with quadratic divergences was raised by L. Susskind~\cite{susskind}.

Most calculations in the Standard Model (including some models with SUSY) are performed using the dimensional regularization  and very successful. Although it is well-known that the dimensional regularization has certain complications with the definition of $\gamma_{5}$, and thus the treatment of the chiral anomaly, for example, is subtle~\cite{siegel},  it works well for anomaly-free theories.
Nevertheless,  people often say that such a specific dimensional regularization which is free of quadratic divergences cannot be generic.  We here present a view that the dimensional regularization is a variant of the more conventional 
regularization and {\em  generic and physical}.

\section{ Subtractive renormalization}

The quadratic divergence in the mass term is generally renormalized {\em subtractively} in the conventional scheme~\footnote{ It is also called {\em renormalized additively} to distinguish it from the conventional usage of ``subtraction of divergences'' in the multiplicative renormalization}, 
 unlike the {\em multiplicative} renormalization of other divergences  in the Standard Model.
Thus we need to understand some specific features of subtractive renormalization.

\subsection{Mass-independent renormalization}

We start with a re-examination of the mass renormalization of the $\lambda\phi^{4}$ theory defined in Euclidean space with the metric $g_{\mu\nu}=(1,1,1,1)$.

To specify a better defined theory, one may start with  
\begin{eqnarray}
{\cal L}&=&-\frac{1}{2}\phi_{0}(x)[-\Box+m_{B}^{2}](\frac{-\Box+M^{2}}
{M^{2}})^{2}\phi_{0}(x)-\frac{1}{4!}\lambda_{0}\phi_{0}(x)^{4},
\end{eqnarray}
which renders all the Feynman diagrams finite, and renormalize 
the quadratically divergent mass term subtractively
\begin{eqnarray}
m^{2}&=&(m^{2}_{B}-\Delta_{sub}(\lambda_{0},M^{2}))/Z(\lambda_{0},M,m)\nonumber\\
&=&m^{2}_{B}+\lambda_{0}c_{1}M^{2} + .... .
\end{eqnarray}
where $Z$ stands for the wave function renormalization factor and 
$\Delta_{sub}(\lambda_{0},M^{2})$ is the subtraction constant to be defined below in \eqref{2.4}.
An enormous  fine-tuning of the bare mass $m_{B}^{2}$ is required to reproduce the observed mass scale $m^{2}$ in the Standard Model for a cut-off $M^{2}$ which may be of the order of the Planck mass.  This procedure is often called {\em unnatural}.

To implement a {\em specific removal} of the quadratic divergence, which we later argue to be physically sensible and natural, we
next specify the bare Lagrangian by 
\begin{eqnarray}\label{2.4}
{\cal L}&=&-\frac{1}{2}\phi_{0}(x)[-\Box+m_{0}^{2}](\frac{-\Box+M^{2}}
{M^{2}})^{2}\phi_{0}(x)-\frac{1}{4!}\lambda_{0}\phi_{0}(x)^{4}\nonumber
\\
&&+\frac{1}{2}\Delta_{sub}(\lambda_{0},M^{2})\phi_{0}(x)^{2},
\end{eqnarray}
where the free propagator is given by 
\begin{eqnarray}
\int d^{4}xe^{ipx}\langle T \phi_{0}(x)\phi_{0}(0)\rangle=\frac{1}{p^{2}+m_{0}^{2}}(\frac{M^{2}}{p^{2}+M^{2}})^{2},
\end{eqnarray}
and the quantity ,
\begin{eqnarray}\label{2.6}
\Delta_{sub}(\lambda_{0},M^{2})
\end{eqnarray} 
is chosen such that all the induced mass terms proportional to $M^{2}$ are completely subtracted.  Our definition of the {\em bare
mass} $m^{2}_{0}$ in \eqref{2.4} and in the following  equations
differs from the common definition~\cite{zinn-justin}, which is given by
\begin{eqnarray}
m^{2}_{B}=m_{0}^{2}+\Delta_{sub}(\lambda_{0},M^{2}), 
\end{eqnarray}
 but in accord with the bare mass in the 
dimensional regularization.  All the variables are now renormalized multiplicatively
\begin{eqnarray}\label{2.6}
\phi_{0}(x)&=&\sqrt{Z( \lambda_{0}, M, m_{0})}\phi(x),\nonumber\\
\lambda_{0}&=&\frac{Z_{\lambda}(\lambda_{0},  M, m_{0})}{Z^{2}(\lambda_{0},  M, m_{0})}\lambda,\nonumber\\
m^{2}_{0}&=&\frac{Z_{m}(\lambda_{0},  M, m_{0})}{Z(\lambda_{0},  M, m_{0})}m^{2}.
\end{eqnarray}
A crucial feature in this scheme is the choice of the subtraction term that is {\em independent} of $m_{0}^{2}$
\begin{eqnarray}\label{2.8}
m_{0}\frac{d}{dm_{0}}\Delta_{sub}(\lambda_{0},M^{2})=0,
\end{eqnarray}
which implies $\Delta_{sub}(\lambda_{0},M^{2})=M^{2}f(\lambda_{0})$.
To ensure this property, it is important to subtract all the quadratic divergences up to any finite order in perturbation theory {\em  before} any multiplicative renormalization (i.e., bare perturbation theory). 
We here illustrate that the quadratic divergence in the scalar field theory in fact satisfies the condition \eqref{2.8}, since it is crucial for our entire analysis; we examine a direct evaluation of the two-loop mass term which is also discussed in Fig.1b later. 
\begin{center}
\includegraphics[scale=0.4,clip]{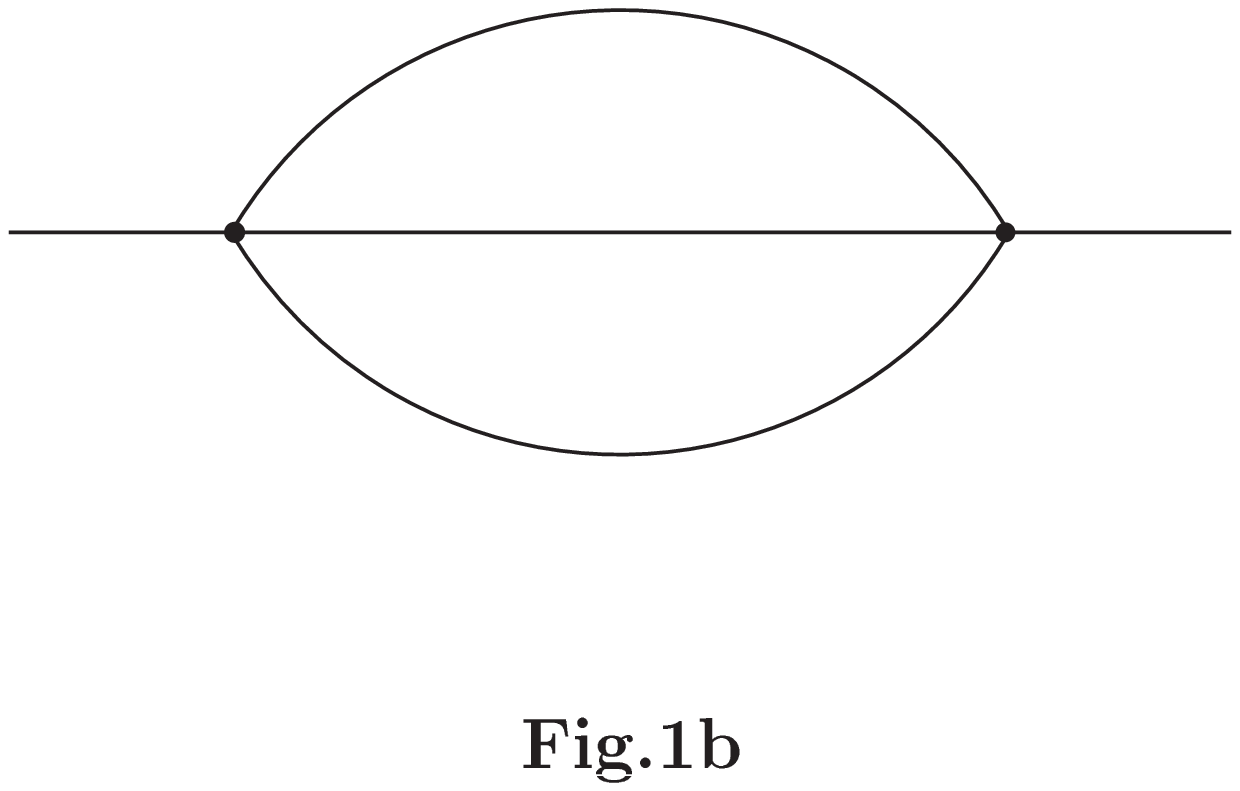}
\end{center}
To analyze the quadratic divergence in this diagram, it is sufficient to analyze the case with vanishing external momentum. We thus examine the integral of the form (without writing 
$\frac{M^{4}}{(k^{2}+M^{2})^{2}}$ here and most cases in the following)
\begin{eqnarray}
\frac{\lambda_{0}^{2}}{3!}\int d^{4}kd^{4}l\frac{1}{k^{2}+m_{0}^{2}}\frac{1}{(k+l)^{2}+m_{0}^{2}}\frac{1}{l^{2}+m_{0}^{2}}
\end{eqnarray}
and confirm that the term proportional to $$\lambda_{0}^{2}M^{2}\ln(M^{2}/m_{0}^{2}),$$ which spoils our assumption \eqref{2.8}, does not appear in the quadratic divergence~\cite{fujikawa}.

To support the general validity of the assumption in \eqref{2.8}, 
which also implies 
\begin{eqnarray}
m\frac{d}{dm}\Delta_{sub}(\lambda_{0},M^{2})=0,
\end{eqnarray}
for the renormalization scheme in \eqref{2.6},
  we here give two arguments. 
If the property \eqref{2.8} is inevitably violated when one subtracts the quadratic divergences in defining general Green's functions, the Callan-Symanzik equation for the $\lambda \phi^{4}$ theory~\cite{callan,symanzik} would contain an extra inhomogeneous term proportional to $M^{2}$ coming from $m_{0}\frac{d}{dm_{0}}\Delta_{sub}$ in the mass insertion term, while the Green's functions are free of quadratic divergences by assumption. The derivation of the Callan-Symanzik equation by a variation of the bare mass $m_{0}^{2}$ in the Lagrangian  \eqref{2.4} using Schwinger's action principle (as is expected in 
the path integral representation $\langle \phi_{0}(x_{1}) ..... \phi_{0}(x_{n})\rangle=\int{\cal D}\phi_{0} \phi_{0}(x_{1}) ..... \phi_{0}(x_{n})\exp[\int d^{4}x{\cal L}]$), 
\begin{eqnarray}
m_{0}\frac{d}{dm_{0}}\langle \phi_{0}(x_{1}) ..... \phi_{0}(x_{n})\rangle=\langle \left(m_{0}\frac{d}{dm_{0}}\int d^{4}x{\cal L}\right) \phi_{0}(x_{1}) ..... \phi_{0}(x_{n})\rangle,
\end{eqnarray}
would then be inconsistent. 
 Secondly, the above subtraction procedure of the quadratic divergence is analogous to the dimensional regularization where the quadratic divergence is completely subtracted
before the conventional multiplicative renormalization; the consistent operation of the dimensional regularization
suggests that the choice $\Delta_{sub}(\lambda_{0},M^{2})$ in \eqref{2.8}
is possible.

The same result  \eqref{2.8} is realized more directly by rewriting the Lagrangian ${\cal L}$ as 
\begin{eqnarray}\label{2.10}
{\cal L}&=&-\frac{1}{2}\phi_{0}(x)[-\Box(\frac{-\Box+M^{2}}
{M^{2}})^{2}]\phi_{0}(x)-\frac{1}{2}m_{0}^{2}\phi_{0}(x)\phi_{0}(x)\nonumber\\
&&-\frac{1}{4!}\lambda_{0}\phi_{0}(x)^{4}
+\frac{1}{2}\Delta_{sub}(\lambda_{0},M^{2})\phi_{0}(x)^{2}
\end{eqnarray}
and treating the mass term as a part of the interaction. 
This {\em mass-independent renormalization scheme}, which is generally understood in the present paper as a prescription to treat  the mass term as an interaction, was introduced to derive the homogeneous renormalization group equation~\cite{weinberg2}.  The formulation of the homogeneous renormalization group equation by 't Hooft~\cite{t hooft2}, which is based on the dimensional regularization, does not encounter the quadratic divergence and one can directly work with the massive scalar theory.  The complication in formulating the mass-independent scheme for a scalar theory due to the quadratically divergent  mass was noted in~\cite{weinberg2}, and
Weinberg analyzed   exclusively the gauge theory consisting of gauge fields and fermions, which is free of quadratic divergences.
The infrared divergence of the massless scalar also complicates the analysis in the present model in \eqref{2.10}~\cite{zinn-justin}.

To cope with infrared divergences, we later operate in a scheme different from the original scheme of Weinberg
 and discuss how to define mass-independent renormalization factors basically in the {\em massive} perturbation theory.
 But for the moment, we discuss the general properties of quadratic divergences on the basis of \eqref{2.10}.
 \\
 \\
 {\bf Analysis of quadratic divergences:}\\

We sketch how the systematic subtraction of the quadratic divergence by  $\Delta_{sub}(\lambda_{0},M^{2})$  works.
This analysis of the quadratic divergence is more transparent in the above mass independent bare perturbation theory \eqref{2.10}. We thus start with the propagator given by 
\begin{eqnarray}\label{2.11}
\int d^{4}xe^{ipx}\langle T \phi_{0}(x)\phi_{0}(0)\rangle=\frac{1}{p^{2}}(\frac{M^{2}}{p^{2}+M^{2}})^{2}.
\end{eqnarray}
We first note that the {\em "primitive" quadratically divergent (single-particle irreducible) diagram which does not contain any quadratically divergent sub-diagrams is infrared finite}.
 Here the quadratically divergent diagrams mean the diagrams 
whose superficial degree of divergence is two, and infrared finiteness means that there arises no singularity for $m_{0}^{2}\rightarrow 0$ in  quadratically divergent terms.
Some of the examples are given in Fig.1.

\begin{center}
\includegraphics[scale=0.4,clip]{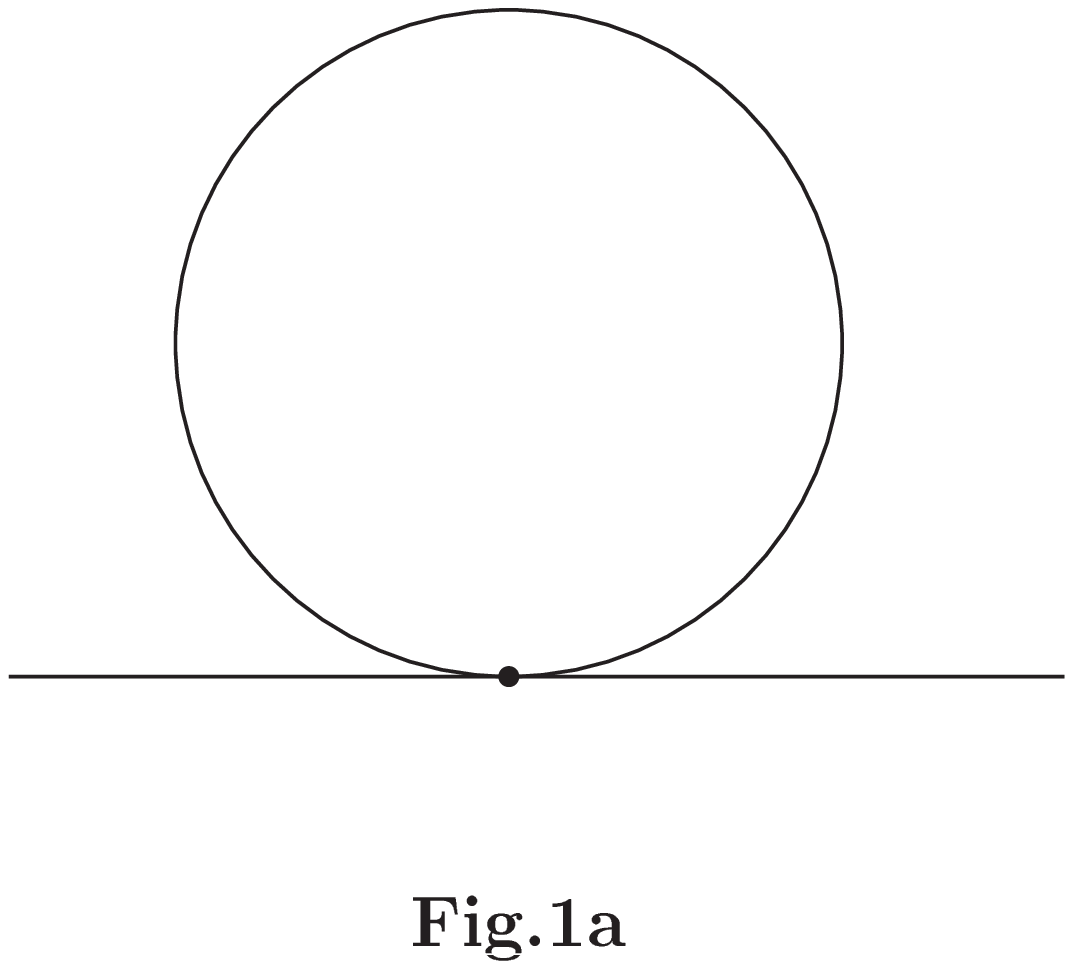}
\hspace{3mm}
\includegraphics[scale=0.4,clip]{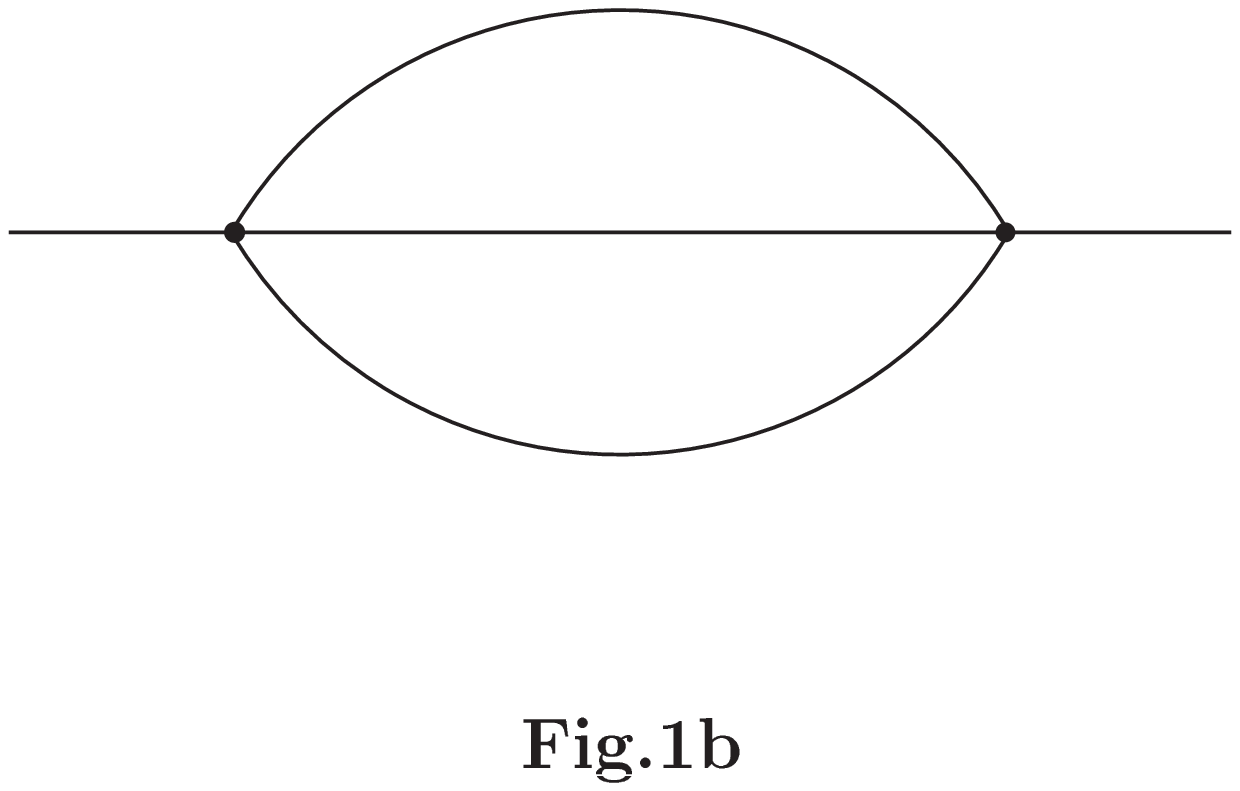}
\end{center}

\vspace{2mm}

\begin{center}
\includegraphics[scale=0.4,clip]{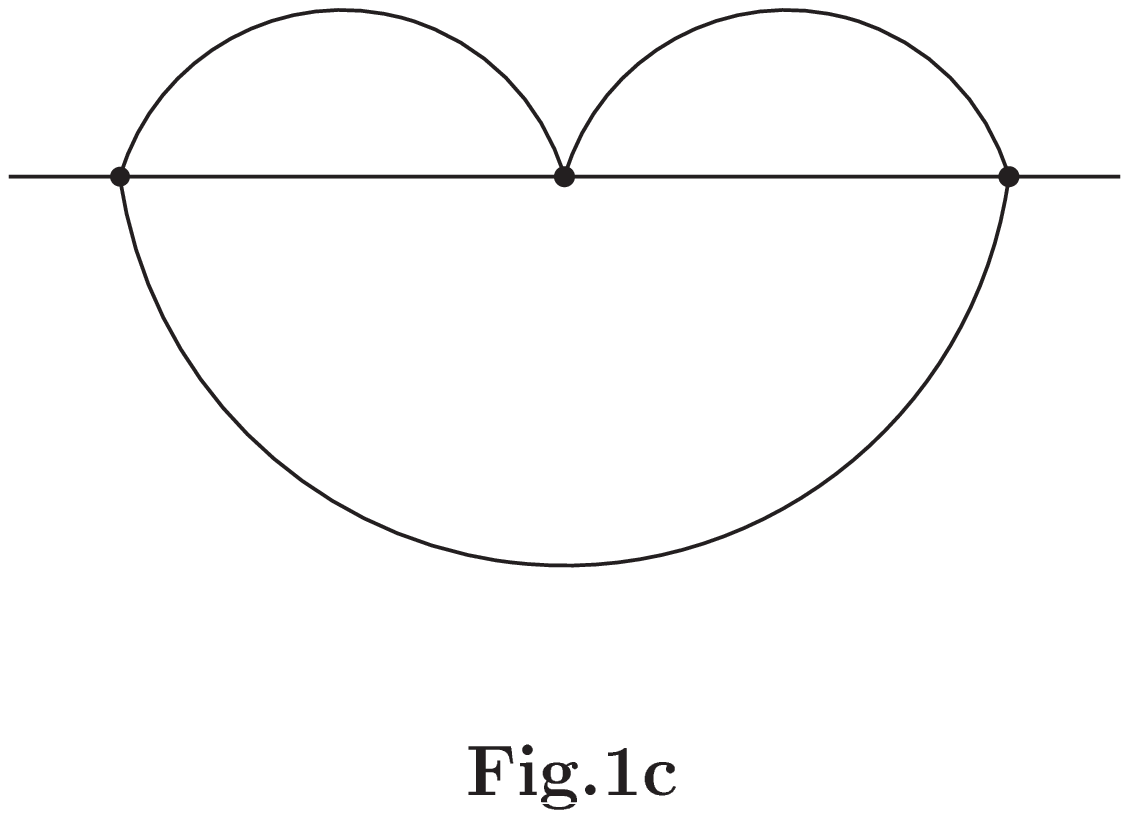}
\hspace{3mm}
\includegraphics[scale=0.4,clip]{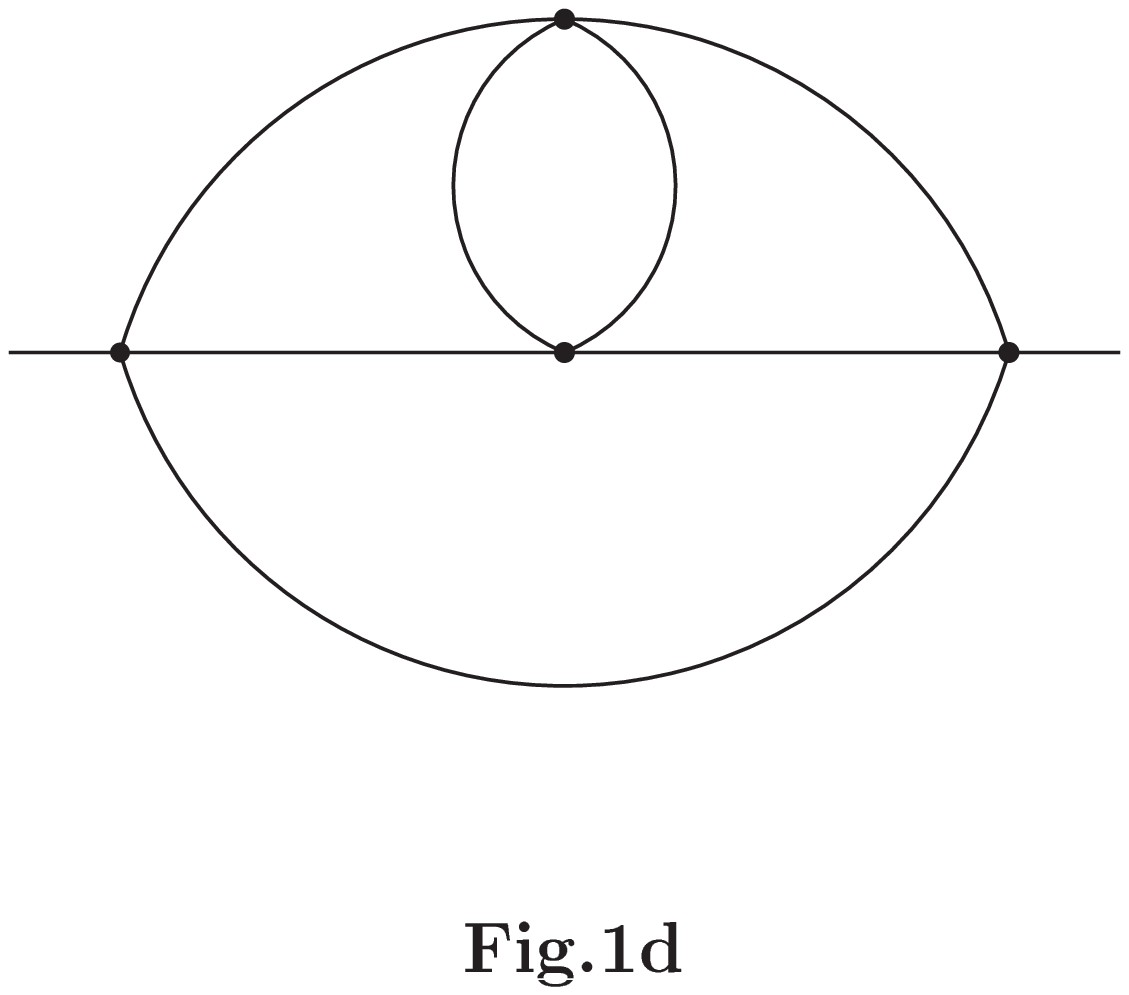}
\end{center}

Fig.1a has been already discussed. The infrared 
finiteness of Fig.1b has been already mentioned. One can confirm the infrared finiteness of Figs.1c and 1d by a power counting argument.   
One may then evaluate any of these self-energy diagrams in Fig.1 to obtain
\begin{eqnarray}
\Sigma(p^{2}, M^{2},\lambda_{0})
\end{eqnarray}
 in the mass independent scheme using the propagator \eqref{2.11}. 
 
 We then subtract the quadratic divergence by  
\begin{eqnarray}\label{2.13}
\tilde{\Sigma}(p^{2}, M^{2},\lambda_{0})&=&\Sigma(p^{2}, M^{2},\lambda_{0})-\Sigma(0, M^{2},\lambda_{0})\nonumber\\
&\equiv&p^{2}A(p^{2}/M^{2},\lambda_{0}).
\end{eqnarray}
The constant $\Sigma(0, M^{2},\lambda_{0})$ constitutes a part of the counter term  $\Delta_{sub}(\lambda_{0},M^{2})$ in \eqref{2.6} in the corresponding order in  perturbation theory.
 
  The quantity $\tilde{\Sigma}(p^{2},M^{2},\lambda_{0})$ identically vanishes for  
 {\em massless tadpole-type diagrams} such as Fig.1a, and thus our prescription agrees with the dimensional regularization. But the {\em massive tadpole-type diagrams} are not eliminated by our prescription 
in accord with the prescription in the dimensional regularization.
We thus understand the specific property of the dimensional regularization where the massless tadpole is completely eliminated but the massive tadpole is not~\cite{t hooft1, collins}.
The quantity $\tilde{\Sigma}(p^{2},M^{2},\lambda_{0})$ thus defined is logarithmically divergent in general in the ultraviolet for large $M$, and $A(p^{2}/M^{2},\lambda_{0})$ in \eqref{2.13} generally contains the (logarithmic) infrared singularity at $p^{2}=0$ .

When one analyzes quadratically divergent diagrams  which contain
one or more {\em "primitive" quadratically divergent sub-diagrams}, one needs to take care of the possible {\em infrared} singularity. Some examples of these diagrams are shown in Fig.2. 
\vspace{2mm}

\begin{center}
\includegraphics[scale=0.6,clip]{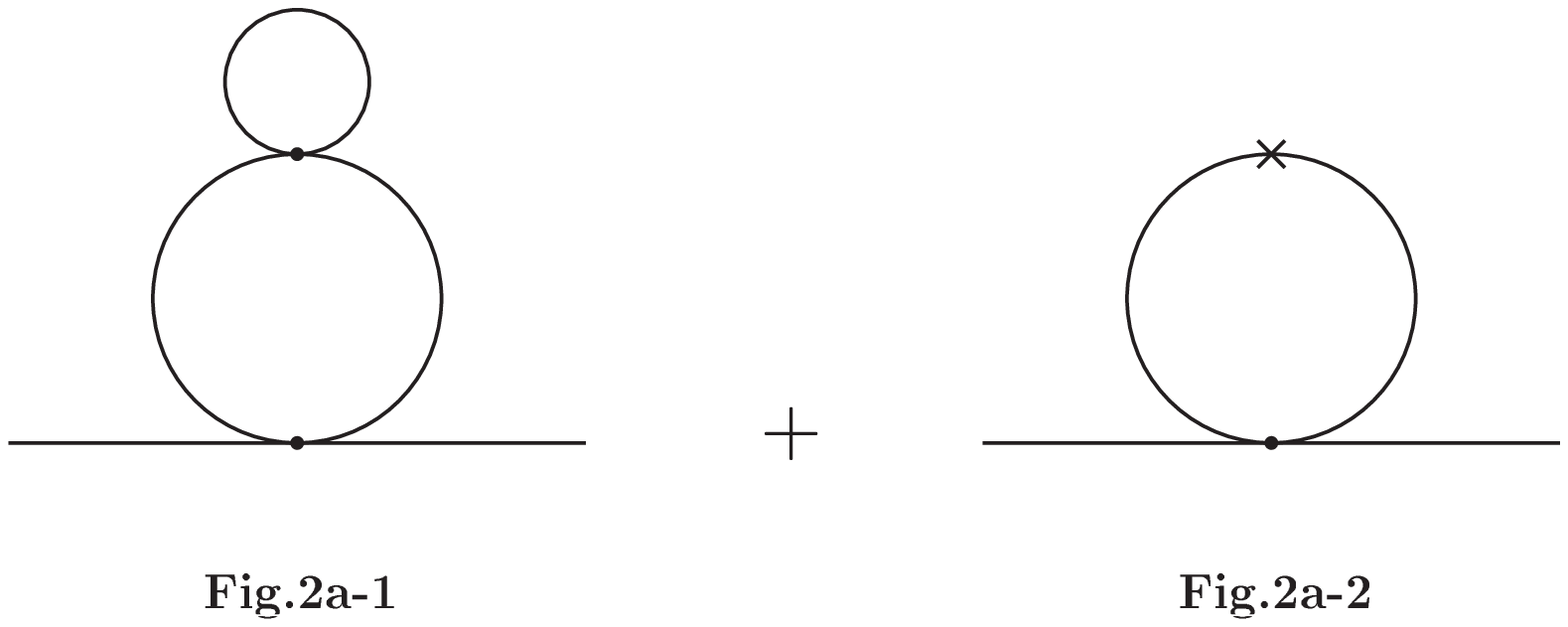}
\end{center}

\vspace{2mm}

\begin{center}
\includegraphics[scale=0.6,clip]{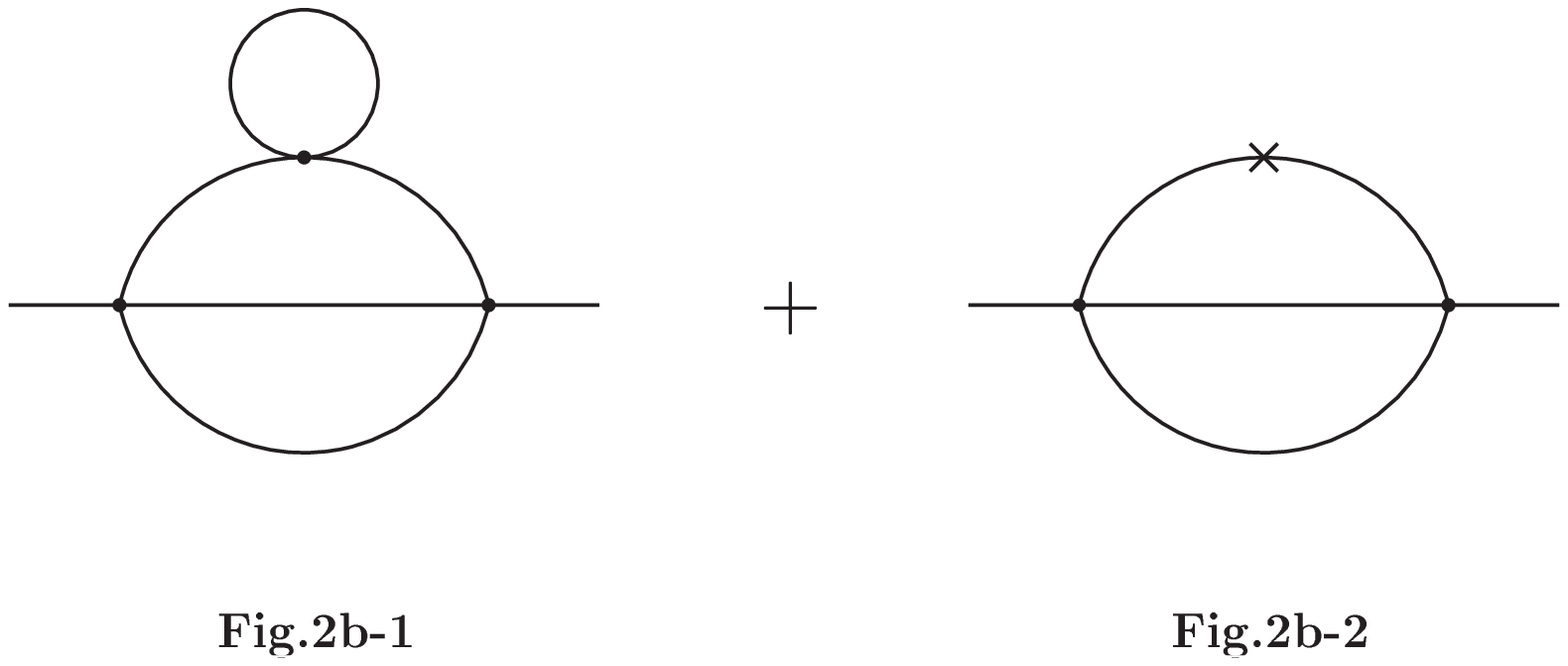}
\end{center}

\vspace{2mm}

\begin{center}
\includegraphics[scale=0.6,clip]{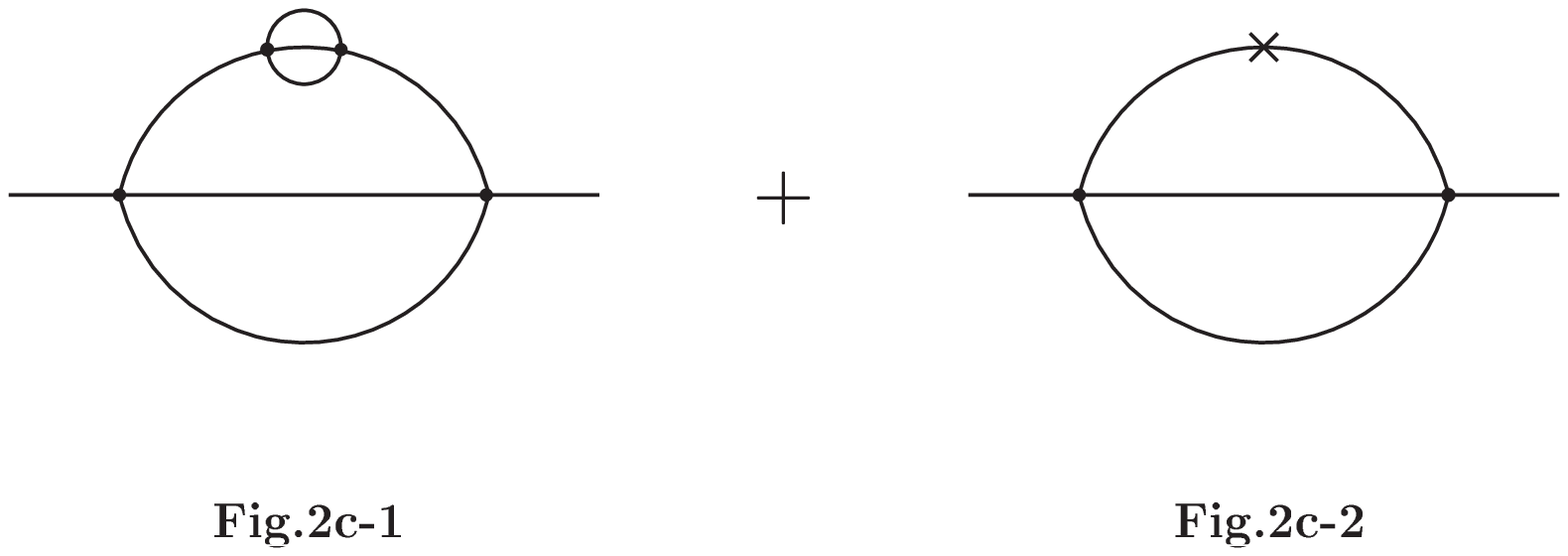}
\end{center}
One can confirm that Figs.2a-1, 2b-1 and 2c-1 are all infrared divergent for the massless propagator, and those infrared divergences are not controlled by the external Euclidean momentum flowing into the diagrams.
But when one combines Fig.2a-1 with  Fig.2a-2, for example, the infrared divergence
is cancelled. Here the cross in Fig.2a-2 stands for $-\Sigma(0, M^{2},\lambda_{0})$ in \eqref{2.13} corresponding to Fig.1a.

Similarly, the combinations of 
Fig.2b-1 and Fig.2b-2 or Fig.2c-1 and Fig.2c-2 are infrared finite if one 
uses $-\Sigma(0, M^{2},\lambda_{0})$  in \eqref{2.13} corresponding to Fig.1a or Fig.1b,
respectively.

 In those combinations, one effectively replaces $\Sigma(p^{2}, M^{2},\lambda_{0})$ by $\tilde{\Sigma}(p^{2}, M^{2},\lambda_{0})$
as in \eqref{2.13} for self-energy sub-diagrams and thus the diagrams with {\em massless tadpole} insertions, such as Fig.2a and Fig.2b, are completely eliminated
in accord with the dimensional regularization. 
One then defines the over-all subtraction constants of quadratic divergences by setting $p^{2}=0$ in those (surviving) {\em infrared-free combinations} in Fig.2. The subtraction constant $\Delta_{sub}(\lambda_{0},M^{2})$  is given by the sum of all these subtraction constants in each given order in perturbation theory.

This subtraction of quadratic divergences works also for the {\em massive} perturbation theory (as is explained in detail later). The present prescription is thus close to that in dimensional regularization, and in fact one may regard our prescription as a  {\em Lagrangian implementation of  dimensional 
regularization} when it comes to the elimination of quadratic divergences.

By this procedure, one can generate the self-energy amplitudes order by order in the bare perturbation theory which are free of quadratic divergences.
One then applies the general multiplicative renormalization procedure in the bare perturbation theory to those self-energy amplitudes at the off-shell point $p^{2}=\mu^{2}$ to define the wave function renormalization factor; the quantity in \eqref{2.13}, $$\tilde{\Sigma}(p^{2}, M^{2},\lambda_{0})$$ generally contains both coupling constant and wave function renormalization factors.

The mass insertion diagrams or the four-point proper vertices, which do not directly induce the quadratic divergence, are handled after removing the possible quadratic divergences in sub-diagrams by the procedure described above. 
In practice, however, one needs a careful treatment of infrared singularities in {\em those logarithmically (ultraviolet) divergent diagrams}. In particular, mass insertion diagrams contain infrared divergences which are not controlled by external momentum flowing into the diagrams.
 See, for example, mass insertion diagrams in Fig.3 which are infrared divergent.

\begin{center}
\includegraphics[clip,scale=0.45]{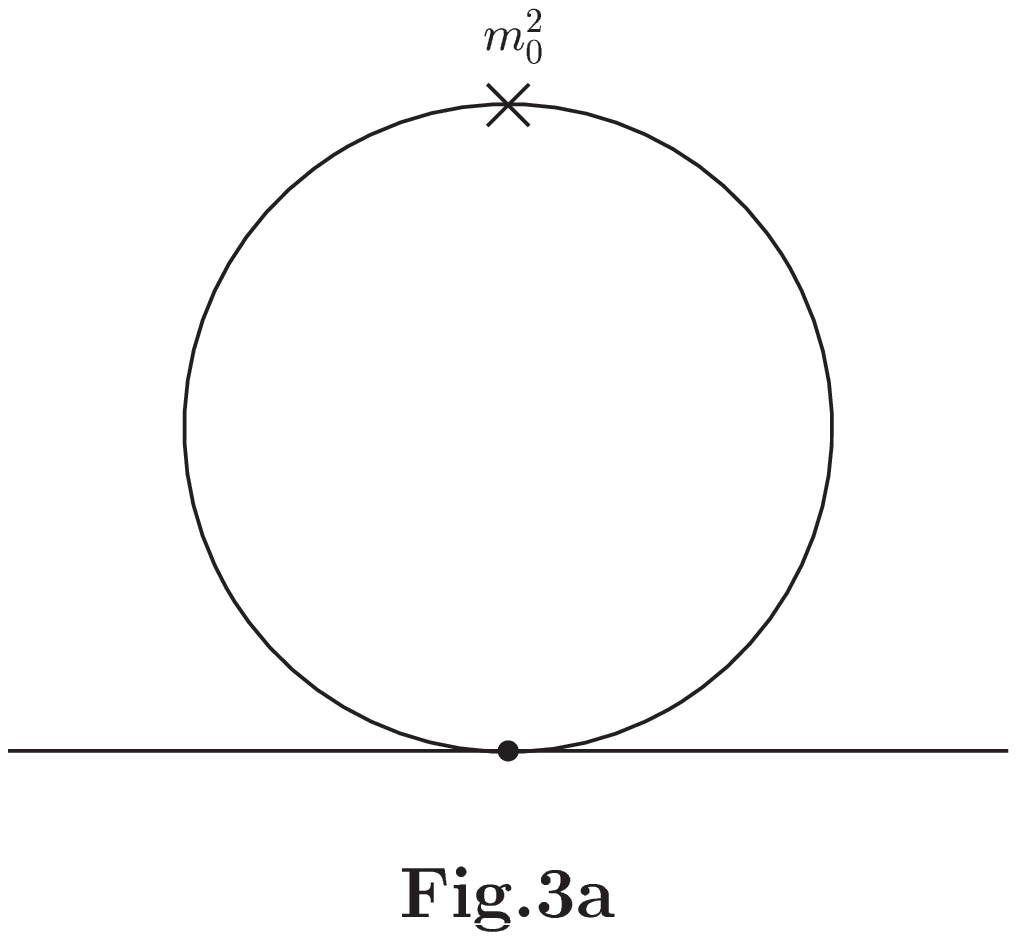}
\hspace{5mm}
\includegraphics[clip,scale=0.45]{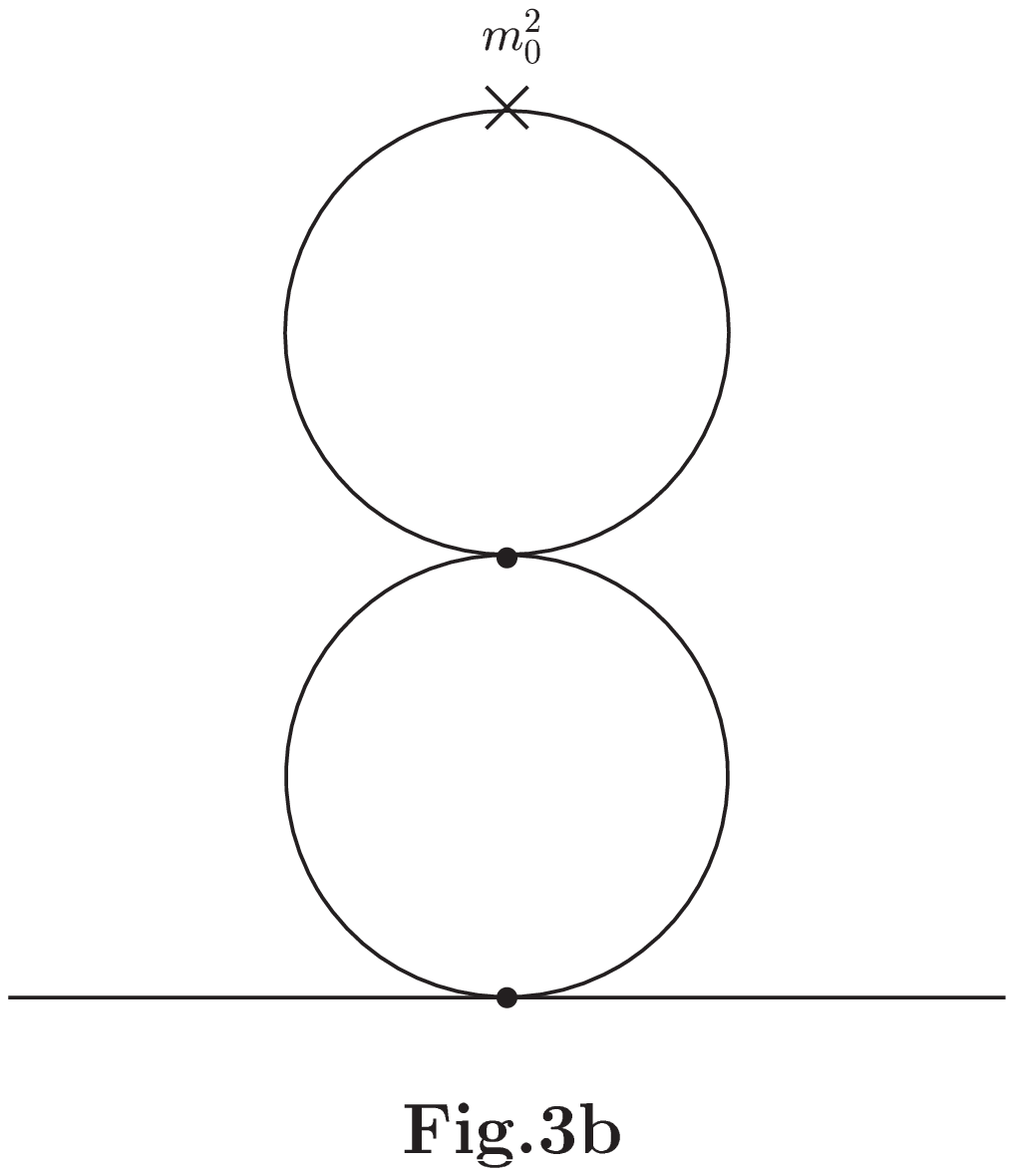}
\hspace{5mm}
\end{center}

\vspace{3mm}

\begin{center}
\includegraphics[clip,scale=0.45]{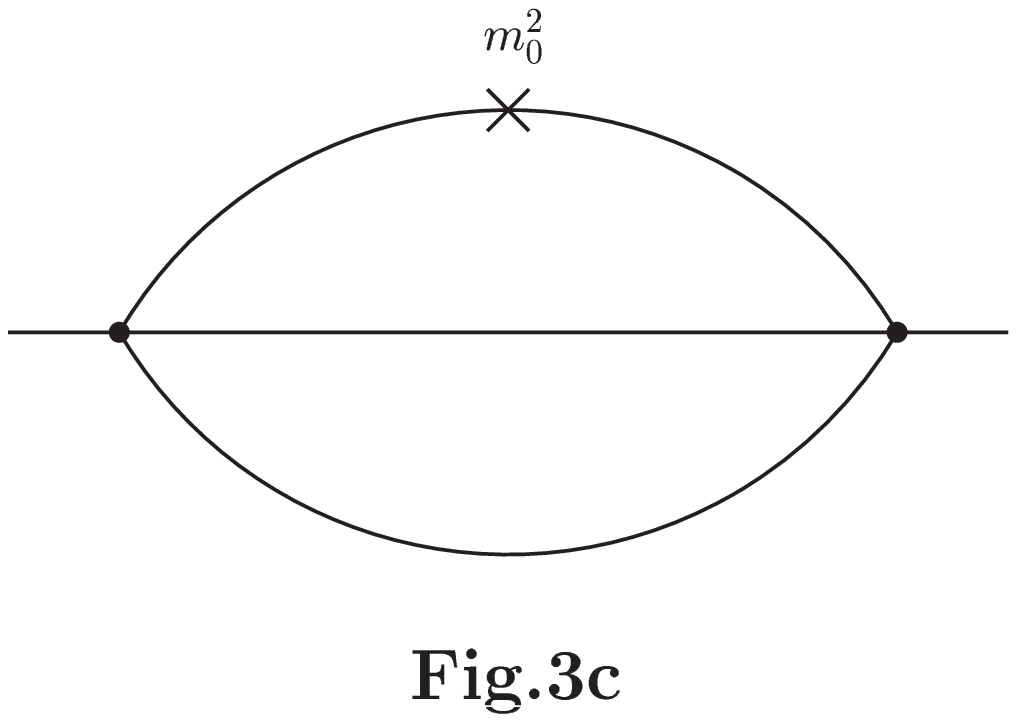}
\end{center}

\noindent {\bf Analysis of infrared divergences:}\\

One needs to take care of the infrared divergence in those 
diagrams such as in Fig.3. To deal with the infrared divergence in a systematic manner we define a  massive perturbation theory defined by~\cite{fujikawa}
\begin{eqnarray}\label{2.14}
{\cal L}&=&-\frac{1}{2}\phi_{0}(x)[-\Box+\mu^{2}](\frac{-\Box+M^{2}}
{M^{2}})^{2}\phi_{0}(x)-\frac{1}{4!}\lambda_{0}\phi_{0}(x)^{4}\nonumber
\\
&&-\frac{1}{2}(m_{0}^{2}-\mu^{2})\phi_{0}(x)(\frac{-\Box+M^{2}}{M^{2}})^{2}\phi_{0}(x)\nonumber\\
&&+\frac{1}{2}\Delta_{sub}(\lambda_{0},M^{2})\phi_{0}(x)^{2},
\end{eqnarray}
which is manifestly free of infrared divergences for the perturbation theory using the propagator defined by the first term in \eqref{2.14} for $\mu\neq 0$, 
\begin{eqnarray}\label{2.15}
\int d^{4}xe^{ipx}\langle T \phi_{0}(x)\phi_{0}(0)\rangle=\frac{1}{p^{2}+\mu^{2}}(\frac{M^{2}}{p^{2}+M^{2}})^{2},
\end{eqnarray}
although the isolation of the quadratic divergence is more transparent in the mass independent scheme \eqref{2.10}.  Note that \eqref{2.14} is invariant under the change of $\mu$. We here depart from the original formulation of Weinberg~\cite{weinberg2} and define the mass independent renormalization factors in the generic form 
\begin{eqnarray}\label{2.17}
&&\phi_{0}(x)=\sqrt{Z( \lambda_{0}, m_{0}, M, \mu)}\phi(x),\nonumber\\
&&m_{0}^{2}=\frac{Z_{m}(\lambda_{0}, m_{0},  M, \mu)}{Z(\lambda_{0},  m_{0},  M, \mu)}m^{2},\nonumber\\
&&\lambda_{0}=\frac{Z_{\lambda}(\lambda_{0},  m_{0},  M, \mu)}{Z^{2}(\lambda_{0},  m_{0},  M, \mu)}\lambda 
\end{eqnarray}
for the {\em massive theory} with $\mu\neq 0$.

We first recall the relation
\begin{eqnarray}\label{2.18}
\frac{1}{p^{2}+m^{2}_{0}}&=&\frac{1}{p^{2}}+\frac{1}{p^{2}}(-m^{2}_{0})\frac{1}{p^{2}}+
\frac{1}{p^{2}}(-m^{2}_{0})\frac{1}{p^{2}}(-m^{2}_{0})\frac{1}{p^{2}}+ ...\nonumber\\
&=&\frac{1}{p^{2}+\mu^{2}+ m^{2}_{0}-\mu^{2}}\nonumber\\
&=&\frac{1}{p^{2}+\mu^{2}}+\frac{1}{p^{2}+\mu^{2}}(-m^{2}_{0}+\mu^{2})\frac{1}{p^{2}+\mu^{2}}\\
&+&\frac{1}{p^{2}+\mu^{2}}(-m^{2}_{0}+\mu^{2})\frac{1}{p^{2}+\mu^{2}}(-m^{2}_{0}+\mu^{2})\frac{1}{p^{2}+\mu^{2}}+...\nonumber
\end{eqnarray}
where the right-hand side of the first line corresponds to the mass-independent scheme
of Weinberg, while the last line corresponds to the present scheme.  To teat the logarithmically divergent diagrams, it is convenient to rewrite this relation as
\begin{eqnarray}\label{log}
\frac{1}{p^{2}+m^{2}_{0}}=\frac{1}{p^{2}+\mu^{2}}-\frac{m_{0}^{2}-\mu^{2}}{(p^{2}+m^{2}_{0})(p^{2}+\mu^{2})} 
\end{eqnarray}
where the second term does not contribute directly to renormalization, while one may use the relation 
\begin{eqnarray}\label{2.19}
\frac{1}{p^{2}+m^{2}_{0}}&=&\frac{1}{p^{2}}-\frac{1}{p^{2}+\mu^{2}}m_{0}^{2}\frac{1}{p^{2}+\mu^{2}}\nonumber\\
&&+\frac{m_{o}^{2}(m_{0}^{2}-2\mu^{2})}{(p^{2}+m^{2}_{0})(p^{2}+\mu^{2})^{2}} 
- \frac{m_{0}^{2}\mu^{4}}{p^{2}(p^{2}+m^{2}_{0})(p^{2}+\mu^{2})^{2}}
\end{eqnarray}
in any quadratically divergent diagrams; the last two terms do not contribute directly to renormalization. These relations are valid with the regulator $(\frac{M^{2}}{p^{2}+M^{2}})^{2}$ added.

For example, the direct evaluation of Fig.1a gives 
\begin{eqnarray}\label{2.21}
\frac{\lambda_{0}}{2}\int \frac{d^{4}p}{(2\pi)^{4}}\frac{1}{p^{2}+m^{2}_{0}}&=&
\frac{\lambda_{0}}{32 \pi^{2}}[M^{2}-m_{0}^{2}\ln \frac{M^{2}}{m_{0}^{2}}]\nonumber\\
&=&\frac{\lambda_{0}}{32 \pi^{2}}[M^{2}-m_{0}^{2}\left(\ln \frac{M^{2}}{\mu^{2}}+\ln \frac{\mu^{2}}{m_{0}^{2}}\right)]
\end{eqnarray}
which agrees with the result of the dimensional regularization (1.13) if one subtracts the quadratic 
divergence $M^{2}$ following our prescription. On the other hand, our \eqref{2.19}
gives
\begin{eqnarray}\label{2.22}
&&\frac{\lambda_{0}}{2}\int \frac{d^{4}p}{(2\pi)^{4}}[\frac{1}{p^{2}}-\frac{1}{p^{2}+\mu^{2}}m_{0}^{2}\frac{1}{p^{2}+\mu^{2}}]\nonumber\\
&=&\frac{\lambda_{0}}{32 \pi^{2}}[M^{2}-m_{0}^{2}\left(\ln \frac{M^{2}}{\mu^{2}}-1\right)]
\end{eqnarray}
and thus the divergent parts are accurately given without encountering infrared divergences; the result \eqref{2.21} is recovered if one recalls the $\mu$ independence and that $\mu^{2}=m_{0}^{2}$ reproduces the exact result in \eqref{2.18}, or by a direct evaluation of the rest of terms in \eqref{2.19}.

As for Fig.2a-1, which contains the (formally) overlapping mass and coupling constant renormalizations, the direct evaluation with our prescription of the elimination of the quadratic divergence
gives
\begin{eqnarray}\label{2.23}
&&\frac{\lambda_{0}}{2}\int \frac{d^{4}p}{(2\pi)^{4}}\frac{1}{p^{2}+m^{2}_{0}}\frac{\lambda_{0}}{2}\int \frac{d^{4}q}{(2\pi)^{4}}\frac{1}{(q^{2}+m^{2}_{0})^{2}}\nonumber\\
&=&\frac{\lambda_{0}}{32 \pi^{2}}[-m_{0}^{2}\left(\ln \frac{M^{2}}{\mu^{2}}+\ln \frac{\mu^{2}}{m_{0}^{2}}\right)]
\frac{\lambda_{0}}{32 \pi^{2}}\left(\ln \frac{M^{2}}{\mu^{2}}+\ln \frac{\mu^{2}}{m_{0}^{2}}-1\right)
\end{eqnarray}
while our scheme gives 
\begin{eqnarray}\label{2.24}
&&\frac{\lambda_{0}}{2}\int \frac{d^{4}p}{(2\pi)^{4}}[\frac{1}{p^{2}}-\frac{1}{p^{2}+\mu^{2}}m_{0}^{2}\frac{1}{p^{2}+\mu^{2}}]\frac{\lambda_{0}}{2}\int \frac{d^{4}q}{(2\pi)^{4}}\frac{1}{(q^{2}+\mu^{2})^{2}}\nonumber\\
&=&\frac{\lambda_{0}}{32 \pi^{2}}[-m_{0}^{2}\left(\ln \frac{M^{2}}{\mu^{2}}-1\right)]
\frac{\lambda_{0}}{32 \pi^{2}}\left(\ln \frac{M^{2}}{\mu^{2}}-1\right)
\end{eqnarray}
and the divergent parts are correctly represented without encountering infrared divergences; again the exact result \eqref{2.23} is reproduced by the $\mu$-independence argument, or by direct evaluations of the rest of terms in 
\eqref{log} and \eqref{2.19}.

As another example, we examine the two-loop self-energy diagram in Fig.1b using the main parts of \eqref{2.19},
\begin{eqnarray}
&&\frac{\lambda^{2}_{0}}{3!}\int \frac{d^{4}l}{(2\pi)^{4}}\int \frac{d^{4}q}{(2\pi)^{4}}[\frac{1}{l^{2}}-\frac{1}{l^{2}+\mu^{2}}m_{0}^{2}\frac{1}{l^{2}+\mu^{2}}][\frac{1}{q^{2}}-\frac{1}{q^{2}+\mu^{2}}m_{0}^{2}\frac{1}{q^{2}+\mu^{2}}]\nonumber\\
&&\times[\frac{1}{(l+q+p)^{2}}-\frac{1}{(l+q+p)^{2}+\mu^{2}}m_{0}^{2}\frac{1}{(l+q+p)^{2}+\mu^{2}}]
\end{eqnarray}
where $p$ stands for the external momentum. As far as main divergent parts are concerned, this is equivalent to 
\begin{eqnarray}
&&\frac{\lambda^{2}_{0}}{3!}\int \frac{d^{4}l}{(2\pi)^{4}}\int \frac{d^{4}q}{(2\pi)^{4}}[\frac{1}{(l+q+p)^{2}}\frac{1}{l^{2}}\frac{1}{q^{2}}\nonumber\\
&&-3\frac{1}{(l+q+p)^{2}+\mu^{2}}m_{0}^{2}\frac{1}{(l+q+p)^{2}+\mu^{2}}\frac{1}{l^{2}}\frac{1}{q^{2}}]
\end{eqnarray}
which are {\em infrared finite}. Our  subtraction prescription of the quadratic divergence corresponds to
the subtraction of 
\begin{eqnarray}
&&\frac{\lambda^{2}_{0}}{3!}\int \frac{d^{4}l}{(2\pi)^{4}}\int \frac{d^{4}q}{(2\pi)^{4}}[\frac{1}{(l+q)^{2}}\frac{1}{l^{2}}\frac{1}{q^{2}}]
\end{eqnarray}
and thus the amplitude is reduced to
\begin{eqnarray}\label{2.26}
&&\frac{\lambda^{2}_{0}}{3!}\int \frac{d^{4}l}{(2\pi)^{4}}\int \frac{d^{4}q}{(2\pi)^{4}}[\frac{1}{(l+q)^{2}}\frac{-p^{2}-2p(l+q)}{(l+q+p)^{2}}\frac{1}{l^{2}}\frac{1}{q^{2}}\nonumber\\
&&-3\frac{1}{(l+q+p)^{2}+\mu^{2}}m_{0}^{2}\frac{1}{(l+q+p)^{2}+\mu^{2}}\frac{1}{l^{2}}\frac{1}{q^{2}}].
\end{eqnarray}
  The conventional overlapping divergence does not appear in this evaluation, although logarithmically divergent sub-diagrams (vertex corrections) are included in the second term in accord with \eqref{2.24}.  To supply the renormalization point $\mu$ explicitly in the expression in \eqref{2.26}, one may replace all the massless propagators by, for example,
\begin{eqnarray}
\frac{1}{q^{2}}=\frac{1}{q^{2}+\mu^{2}}+\frac{\mu^{2}}{q^{2}(q^{2}+\mu^{2})}
\end{eqnarray} 
where the last term may be combined with the rest of terms in \eqref{2.19} which give non-leading terms. The first term in \eqref{2.26} gives a wave function renormalization proportional to
$p^{2}\lambda^{2}_{0}\ln\frac{M^{2}}{\mu^{2}}$ and the second term gives a mass renormalization proportional to $m_{0}^{2}\lambda^{2}_{0}(\ln\frac{M^{2}}{\mu^{2}})^{2}$.

As for the coupling constant renormalization, it is logarithmically divergent in the ultraviolet after the elimination of possible quadratic divergences in subdiagrams in the present prescription. For those logarithmically divergent diagrams, the above renormalization constant in \eqref{2.17} is obtained
when renormalized at the vanishing momenta.
For example, in the one-loop level we have (and two more similar terms) using \eqref{log}
\begin{eqnarray}
&&\frac{\lambda_{0}^{2}}{2}\int \frac{d^{4}l}{(2\pi)^{4}}\frac{1}{(l+p)^{2}+\mu^{2}}(\frac{M^{2}}{(l+p)^{2}+M^{2}})^{2}\frac{1}{l^{2}+\mu^{2}}(\frac{M^{2}}{l^{2}+M^{2}})^{2}\nonumber\\
\end{eqnarray}
where $p$ stands for the external momentum.

We have illustrated that our prescription on the basis of a conventional regularization reproduces the main features of the dimensional regularization as to the treatment of quadratic divergences without encountering infrared divergences.  Although this heuristic argument does not constitute a proof, we believe that an extension of our analysis to higher order diagrams is possible.

\section{ Summary and conclusion}

We have illustrated  that the elimination of the quadratic divergence and the mass-independent multiplicative renormalization
\begin{eqnarray} 
&&\phi_{0}(x)=\sqrt{Z( \lambda_{0}, m_{0},  M, \mu)}\phi(x),\nonumber\\
&&m_{0}^{2}=\frac{Z_{m}(\lambda_{0}, m_{0},  M, \mu)}{Z(\lambda_{0},  m_{0},  M, \mu)}m^{2},\nonumber\\
&&\lambda_{0}=\frac{Z_{\lambda}(\lambda_{0},  m_{0},  M, \mu)}{Z^{2}(\lambda_{0},  m_{0},  M, \mu)}\lambda,
\end{eqnarray}
is implemented by the Lagrangian
\begin{eqnarray} \label{3.2}
{\cal L}&=&-\frac{1}{2}\phi_{0}(x)[-\Box+\mu^{2}](\frac{-\Box+M^{2}}
{M^{2}})^{2}\phi_{0}(x)-\frac{1}{4!}\lambda_{0}\phi_{0}(x)^{4}\nonumber
\\
&&-\frac{1}{2}(m_{0}^{2}-\mu^{2})\phi_{0}(x)(\frac{-\Box+M^{2}}{M^{2}})^{2}\phi_{0}(x)\nonumber\\
&&+\frac{1}{2}\Delta_{sub}(\lambda_{0},M^{2})\phi_{0}(x)^{2},
\end{eqnarray}
which is independent of the change in $\mu$~\cite{fujikawa}.  Note that the multiplicative renormalization means that the renormalized mass $m^{2}$ is made small by choosing the bare mass $m_{0}^{2}$ sufficiently small. One defines the perturbation theory with the free propagator 
\begin{eqnarray}
\int d^{4}xe^{ipx}\langle T \phi_{0}(x)\phi_{0}(0)\rangle=\frac{1}{p^{2}+\mu^{2}}(\frac{M^{2}}{p^{2}+M^{2}})^{2},
\end{eqnarray}
which renders all the Feynman diagrams finite, and then regarding the 3rd mass term in
\eqref{3.2} as a 
part of interaction, it generalizes the scheme of Weinberg in a manner which is free of infrared divergences for $\mu^{2}\neq 0$. 
When one renormalizes the theory at vanishing momenta, the {\em arbitrary
parameter $\mu$ specifies the renormalization point}.

The logarithmically divergent diagrams (or sub-diagrams) are handled by the above propagator  in (3.3) or using \eqref{log}
without encountering the infrared divergence.

The treatment of  quadratically divergent diagrams is based on  the replacement as in \eqref{2.19},
\begin{eqnarray}
\frac{1}{p^{2}+m^{2}_{0}}&=&\frac{1}{p^{2}}-\frac{1}{p^{2}+\mu^{2}}m_{0}^{2}\frac{1}{p^{2}+\mu^{2}}\nonumber\\
&&+\frac{m_{o}^{2}(m_{0}^{2}-2\mu^{2})}{(p^{2}+m^{2}_{0})(p^{2}+\mu^{2})^{2}} 
- \frac{m_{0}^{2}\mu^{4}}{p^{2}(p^{2}+m^{2}_{0})(p^{2}+\mu^{2})^{2}}
\end{eqnarray}
We then follow the explained procedure to treat the quadratically divergent diagrams  by maintaining infrared finiteness; the first term on the right-hand side gives $m_{0}^{2}$ independent quadratically divergent mass in quadratically divergent diagrams, which is eliminated by $\Delta_{sub}(\lambda_{0},M^{2})$, and the second term gives a logarithmic renormalization of the mass parameter. The last two terms above do not contribute to logarithmic mass renormalization.
 The present replacement scheme is useful to see that the logarithmic renormalization of $m_{0}^{2}$ is sufficient for mass renormalization and that no renormalization of the parameter $\mu^{2}$ takes place, which is necessary if one wants to identify $\mu$ with the renormalization point. 

 We thus recognize that the 
crucial ingredient of our analysis is the elimination of all the quadratic divergences by the counter term $\Delta_{sub}(\lambda_{0},M^{2})$ with
\begin{eqnarray}
m_{0}\frac{d}{dm_{0}}\Delta_{sub}(\lambda_{0},M^{2})=0,
\end{eqnarray}
in the (generalized) mass-independent scheme, which 
reproduces the essence of the dimensional regularization. The basic understanding is that 
the quadratically divergent mass which is independent of the scale change of the physical mass $m_{0}^{2}$ is unphysical; this is the basis of a bottom-up approach starting with a successful low energy theory. 

A salient property of the present procedure is that the  
symmetry of the Lagrangian, namely the $\mu$ independence in \eqref{3.2}, gives the homogeneous renormalization group equation
\begin{eqnarray}\label{3.5}
\mu\frac{d}{d\mu}\langle \phi_{0}(x_{1})\phi_{0}(x_{2})....\phi_{0}(x_{n})\rangle=0,
\end{eqnarray}
which is analogous to the Ward-Takahashi identity, and the renormalization point $\mu$ is introduced in the process of Feynman diagram evaluation, which is analogous to the evaluation in dimensional regularization~\cite{t hooft1}.  One may recall that the parameter $\mu$ is introduced {\em after} the evaluation of Feynman diagrams in the conventional formulation.
 (A renormalization scheme more general than \eqref{3.2} was discussed in~\cite{fujikawa}, but here we adopt the simplest case  \eqref{3.2}.)
 
The homogeneous  renormalization group equation follows from \eqref{3.5} 
\begin{eqnarray}
\{\mu\frac{\partial}{\partial \mu}+\beta\frac{\partial}{\partial\lambda}-\left(\gamma_{m}m^{2}\right)\frac{\partial}{\partial m^{2}}-n\gamma_{\phi}\}\Gamma_{n}(p_{1}, ...., p_{n})=0,
\end{eqnarray}
where we defined renormalization group parameters by the standard manner 
\begin{eqnarray}
&&\gamma_{m}=-\left(\mu\frac{d}{d\mu}m^{2}(\mu)\right)/m^{2}=\left(\mu\frac{d}{d\mu}(\frac{Z_{m}}{Z})\right)(\frac{Z}{Z_{m}}),\nonumber\\
&&\beta=\mu\frac{d}{d\mu}\lambda(\mu),\nonumber\\
&&\gamma_{\phi}=\frac{1}{2}\frac{1}{Z}\mu\frac{d}{d\mu}Z
\end{eqnarray}
where the derivative is taken with fixed bare parameters and $M$.
\\

In conclusion, the success of the Standard Model without obvious SUSY at LHC may indicate that the treatment of 
the quadratic divergence by dimensional regularization is {\em natural}. Physically this means that the quadratic divergence induced by the cut-off, which is independent of the scale change of the bare mass $m_{0}^{2}$, is {\em kinematical and has no definite physical meaning} and thus can be eliminated without modifying physics.   This is consistent with the derivation of the Callan-Symanzik equation without encountering the quadratic divergence by a comparison of two theories with slightly different bare masses. (Diagramatically, this comparison of two theories results in  a mass insertion which reduces ``primitive'' quadratic divergences to logarithmic ones.) 

To support 
the above view, we illustrated how to reproduce the results of dimensional regularization by a conventional regularization combined with the generalized mass-independent renormalization. Of course, the dimensional regularization gives required results in a much simpler and quicker way and thus  practically far more useful~\cite{t hooft2}. 

Once one accepts the view that the quadratically divergent mass, which is independent of the scale change of the bare mass $m_{0}^{2}$, is eliminated without modifying physics in the Standard 
Model, one may attempt to extend the Standard Model to the energy scale of the Planck mass on the basis of renormalization group analyses such as in~\cite{nielsen},
or one may work with high-energy SUSY schemes since the onset of SUSY is now disconnected from the energy scale of the Higgs mass.  

Finally, we mention the works~\cite{bardeen, iso, kawai, bian, kawamura, cherchiglia} where interesting physical issues related to quadratic divergences are analyzed from various points of view.

\subsection*{Acknowledgments}
I thank Henry Tye for asking to what extent the dimensional regularization is generic.

\end{document}